\documentclass{moriond}

\def\Journal#1#2#3#4{{#1} {\bf #2}, #3 (#4)}

\def\PRD{{\em Phys. Rev.} D}

\def\be{\begin{equation}}
\def\ee{\end{equation}}
\def\bea{\begin{eqnarray}}
\def\eea{\end{eqnarray}}

\usepackage{bm}

\newcommand{\Photo}{\includegraphics[height=35mm]{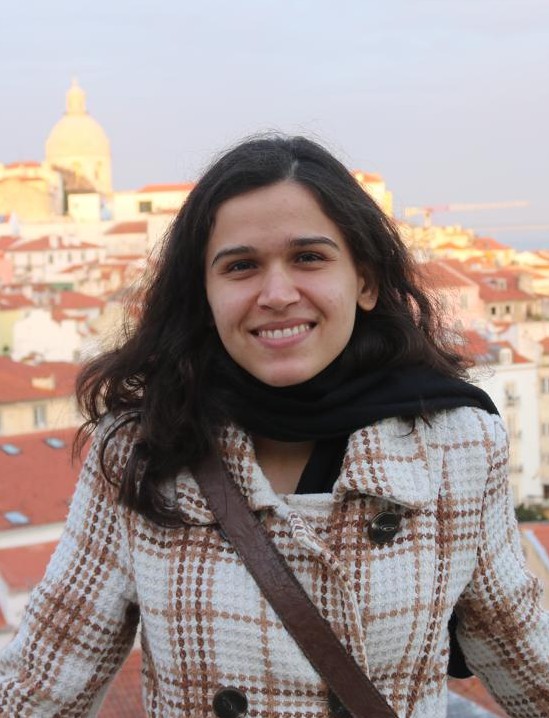}}

\begin{document}
\vspace*{4cm}
\title{Constraining modified gravity with gravitational wave distance measurements}

\author{ ISABELA S. MATOS \footnote{isa@if.ufrj.br}}

\address{Universidade Federal do Rio de Janeiro, Instituto de Física, \\ CEP 21941-972 Rio de Janeiro, RJ, Brazil\\
Départment de Physique Théorique and Center for Astroparticle Physics, Université de Genève, \\ CH-1211 Genève 4, Switzerland}

\maketitle\abstracts{
It has been shown in the literature that detections of gravitational waves (GWs) emitted by binary sources can provide measurements of luminosity distance. The events followed by electromagnetic counterparts are, then, suitable for probing the distance-redshift relation and doing cosmological parameter estimation, as well as investigating modified gravity (MG) models. In the context of effective approaches to MG equivalent to Horndeski, this GW distance differs from the standard electromagnetic luminosity distance due to the presence of  a modified friction in the wave propagation. Here, we investigate how precisely the future-planned interferometer Einstein Telescope will probe such deviations from General Relativity, considering phenomenological parametrizations for both the dark energy equation of state and the GW friction. Despite being an independent test of gravity, for $f(R)$ in particular, we conclude that it may provide weaker constraints than the current available ones.}

\section{Introduction}

\noindent

Since the groundbreaking discovery of LIGO-Virgo in 2015, a lot of effort has been devoted to clarify how much we can learn from gravitational wave (GW) detections in different topics of interest in spacial physics, \textit{e.g.} the equation of state of neutron stars, the current expansion rate of the universe, tests of gravity, compact objects and so on. In particular, the so-called multimesseger events, \textit{i.e.} those that are followed by electromagnetic signals such as $\gamma$-ray bursts, stand out as the ones which allow us to assess more information about the source, \textit{e.g.} its redshift, and also about possible differences in the propagation of gravitons and photons.

In our work \cite{Matos2021} \footnote{in collaboration with M. O. Calvão and I. Waga.} we performed simulations of multimessenger events emitted by binary neutron stars (BNS) for the future planned 3rd generation interferometer Einstein Telescope (ET) \cite{Maggiore2020}. In an optimistic scenario, $10^3$ of such signals are expected to be detected up to redshift 2, which allows us to test modified gravity (MG) theories at cosmological scales in the tensor sector. This can be achieved by parametrizing MG signatures in the evolution, from emittion to detection, of the wave's amplitude in the inspiral phase where perturbation theory applies. Our framework is defined by a perturbed spatially flat FLRW metric,
\begin{equation}
ds^2 = a(\tau)^2\left[ -(1 + 2\Psi)d\tau^2 + (1 - 2\Phi)(\delta_{ij} + h_{ij})dx^idx^j\right]\,,
\end{equation}
where $a$ is the scale factor, $\tau$ is the conformal time, $\Phi$ and $\Psi$ are the usual scalar perturbations and $h_{ij}$ is the transverse traceless GW, specified by the two degrees of freedom $h_+$ and $h_{\times}$.

In the context of cosmology, it has been shown \cite{Bellini2014} that several MG theories can be distinguished at background and linear perturbation levels only through the specification of a tuple
\begin{equation}
\left(\Omega_{m0}, {\mathcal H}, \alpha_M, \alpha_T, \alpha_K, \alpha_B\right)\,,
\end{equation}
where $\Omega_{m0}$ is the matter energy density parameter today, ${\mathcal H}:=a'/a$ is the conformal Hubble parameter $(' := d/d\tau)$ and the $\alpha$'s are time-dependent functions that govern the evolution of perturbations. They all vanish in General Relativity (GR) and their physical meanings are associated with the presence of, respectively, a running Planck mass $M_{\ast}^2$, a tensor speed excess, kineticity and braiding. This setting encompasses the most general theories that give rise to second order field equations with a single scalar field $\phi$, the Horndeski theories. Moreover, the evolution equation for the GW polarization modes is given by
\begin{equation}
h_P'' + (2 + \alpha_M){\mathcal H}h'_P + (1 + \alpha_T)k^2h_P = \Pi_P\,, \label{GW}
\end{equation}
with $P = +, \times$, where $\Pi_P$ is the tensor part of the anisotropic stress. This equation differs from its corresponding one in Einstein's gravity by the introduction of features such as a change in the GW speed of propagation $c_T^2 = 1 + \alpha_T$. However, this was already strongly constrained by the GW170817 event to be equal to the speed of light \cite{Abbot2017}, which motivates restricting to $\alpha_T = 0$.

The particularly important deviation from GR in the propagation of tensor modes is encoded in the friction rate at which they are damped whenever $\alpha_M \neq 0$. This effect gives rise to a new cosmological distance that can be inferred by local interferometers, the GW distance $D^{\mathrm{gw}}$ \cite{Belgacem2019}. It relates to the standard electromagnetic luminosity distance by
\
\begin{equation}
\Xi(z) := \frac{D^{\mathrm{gw}}(z)}{D^{\mathrm{em}}_L(z)} = \exp\left\{\frac{1}{2}\int_0^z \frac{\alpha_M(\tilde{z})}{1 + \tilde{z}}d\tilde{z}\right\} = \sqrt{\frac{M_{\ast}^2(0)}{M_{\ast}^2(z)}}\,. \label{gw_distance}
\end{equation}
In $f(R)$ theories, where our analysis is focused on, this reduces to $\sqrt{1 + f_{R0}}/\sqrt{1 + f_R}$, where $f_R$ is the derivative of the Lagrangian with respect to the Ricci scalar. For viable models that recap GR at high redshifts, this function goes asymptotically to $1 + f_{R0}/2$. Its typical behavior is shown in the right pannel of Fig. \ref{fig:results} for a particular model.

\section{GW simulations}

\noindent

We now summarize the procedure to simulate the GW mock data for the ET with $N_{\mathrm{obs}} = 1000$ events. We considered the waveform emitted at redshift $z$ by a BNS merger up to the 3rd PN correction neglecting spins, which is given by
\begin{equation}
\tilde{h}(f, z, D^{\mathrm{gw}}, \bm{s}) = \frac{\mathcal{A}(f, z, \bm{s})}{D^{\mathrm{gw}}} \left(\frac{c}{f^{7}}\right)^{1/6}e^{i\Phi(f)}\,,
\end{equation}
where $\Phi$ is a phase and $\mathcal{A}$ is a function of frequency, redshift and the parameters of the source $\bm{s} = (m_1, m_2, \iota, \theta, \phi, \psi)$, which stand for, respectively: the masses of the binary components, the angle of orbital inclination, the direction of the line of sight and the GW polarization angle.

Given a gravitational and cosmological model with parameter values $\bm{\Theta}$, we draw the redshift of the i-th binary source from the distribution
\begin{equation}
\rho (z_i| \bm{\Theta}) = N_{z}(\bm{\Theta}) \frac{4\pi [\mathcal{D}_c(z_i, \bm{\Theta})]^2}{(1 + z_i) H(z_i, \bm{\Theta})} r(z_i)\,, \label{z_pdf}
\end{equation}
where $\mathcal{D}_c(z, \bm{\Theta}) = \int_0^z c/H(z, \bm{\Theta})$ is the comoving distance, $N_z$ is a normalization factor and $r(z)$ is the rate of evolution of the BNS mergers.

Accounting for the degeneracy between $D^{\mathrm{gw}}$ and $\iota$ and adding a contribution due to weak lensing, the error in the GW distance, in the Fisher matrix approximation, is given by
\begin{equation}
\sigma^2 = \left[\frac{2 \mathcal{D}^{\mathrm{gw}}(z)}{\mathrm{SNR}(z, \bm{s})}\right]^2 + \left[\sigma_{\mathrm{lens}}(z)\right]^2 \quad \mathrm{with} \quad \mathrm{SNR}^2 = 4 \int_{f_{\mathrm{low}}}^{f_{\mathrm{up}}} \frac{\left|\tilde{h}\right|^2}{S_n}df\,,
\end{equation}
where $S_n$ is the power spectral density of the detector's noise. The GW distances are drawn from a Gaussian distribution with standard deviation $\sigma$ and, therefore, our posterior results in
\begin{equation}
\rho(\bm{\Theta}| \bm{z}, \bm{D}^{\,gw}_{L}, \bm{s} ) \propto \rho (\bm{\Theta}) \exp \Bigg\{-\sum_{i = 1}^{N_{\mathrm{obs}}} \frac{\left[D^{\,gw}_{L,i} - \mathcal{D}^{\,gw}_L(z_i, \bm{\Theta})\right]^2}{2\sigma(z_i, \bm{s}_i)^2}\Bigg\}\,. \label{final_posterior}
\end{equation}

\section{Results}

\noindent

When restricting to $\Lambda$CDM, we find that a thousand GW events followed by electromagnetic counterparts, if detected by ET, would provide measurements of $H_0$ and $\Omega_{m0}$ with accuracies of $\Delta H_0/H_0 \sim 1 \%$ and $\Delta \Omega_{m0}/\Omega_{m0} \sim 7.6 \%$. Secondly, when allowing for a relative difference between GW and luminosity distances in a generic viable $f(R)$ theory, we obtain the bound $|f_{R0}| <  10^{-2}$. In fact, we estimate that $10^{10}$ of such multimessenger events would be needed to reduce this number to $10^{-6}$, which is the current known bound from large scale structure (LSS) and local tests.

\subsection{Parametrizations}

\noindent

\begin{figure}
	\centering
	\vspace{-25pt}
	\hspace{-25pt}
	\includegraphics[scale=0.35]{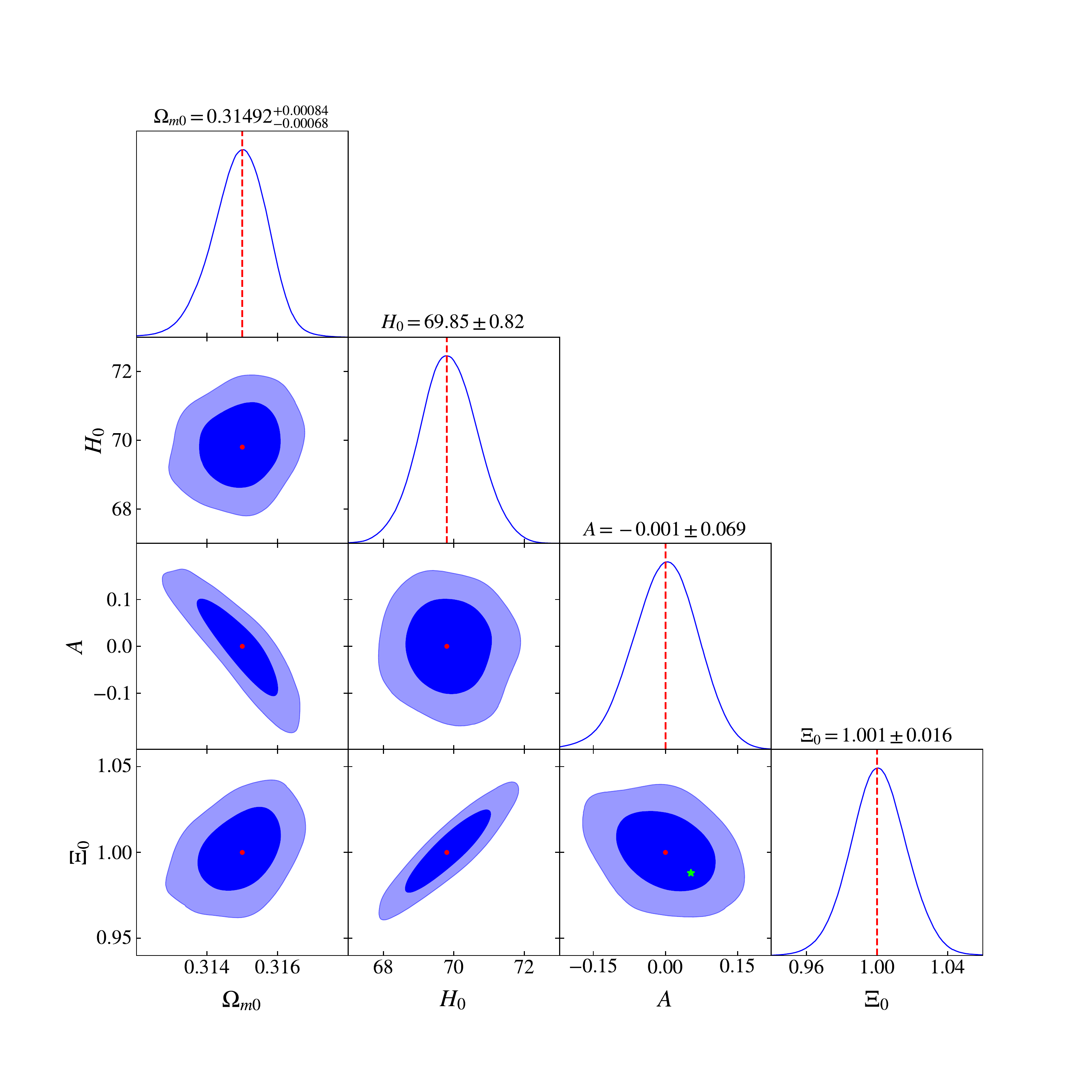}
	\vspace{-20pt}
	\includegraphics[scale=0.4, trim=0 -1.4cm 0 0]{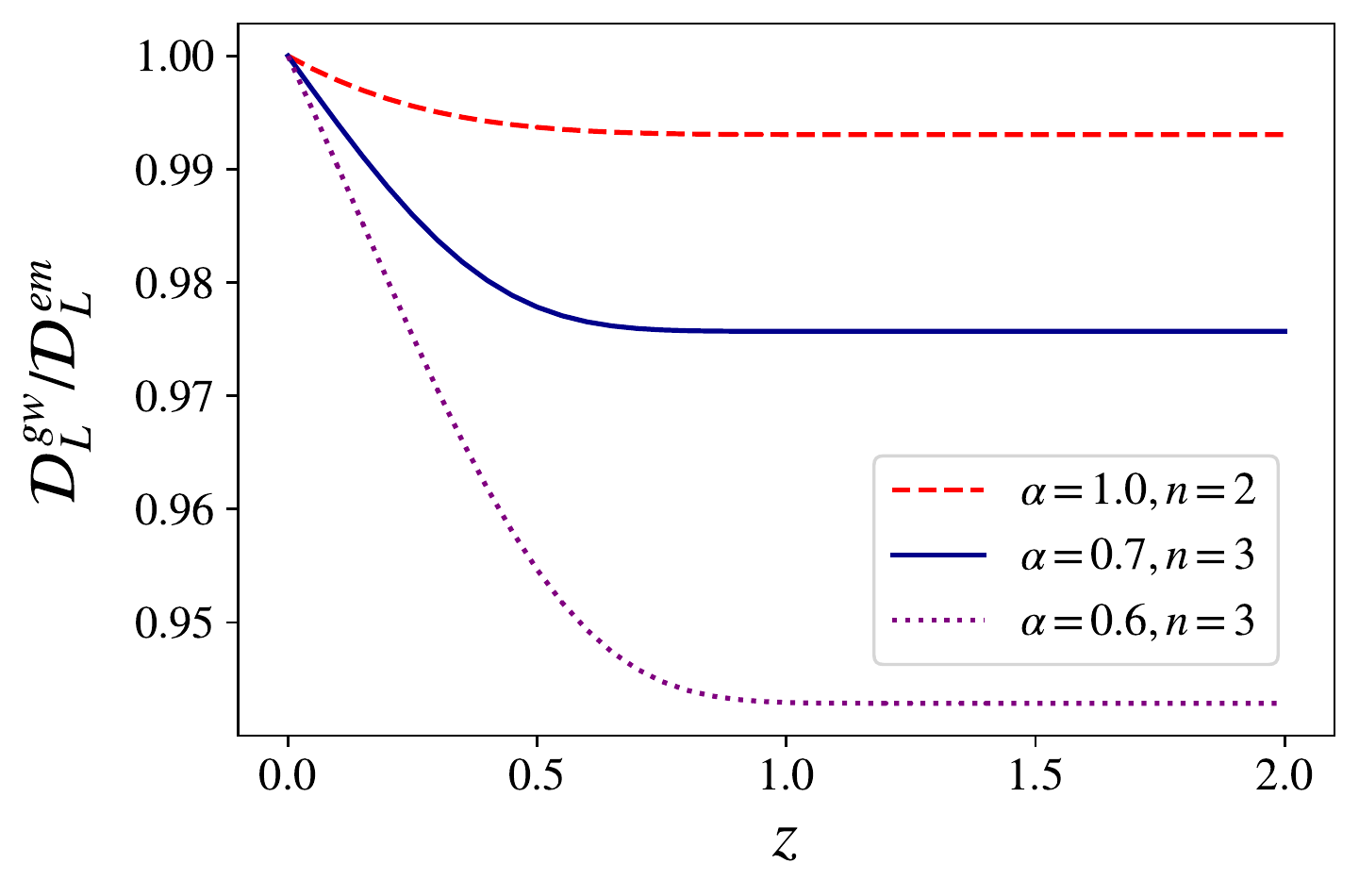}
	\caption{On the left, 68\% and 95\% CLs from simulations of a thousand GW detections assuming the redshifts of the sources are known. Red dots or dashed lines correspond to $\Lambda$CDM. On the right, the theoretical prediction for the ratio of distances in $\gamma$-gravity $f(R)$ for three values of the parameters defining the Lagrangian.}
	\label{fig:results}
\end{figure}

In order to account for MG effects both at the background level and in the friction we propose parametrizations for the dark energy equation of state (EOS) parameter $w_{\mathrm{DE}}$ and the ratio of distances $\Xi$. They are:
\begin{eqnarray}
w_{\mathrm{DE}}(z, A, z_t, z_f) = & -1 - A (z-z_f) (z_t-z) \sin\bigg[\frac{2 \pi z- \pi (z_f+z_t)}{z_t-z_f}\bigg]\,, \quad z_f < z < z_t \\
\Xi (z, \Xi_0, \nu) = & \hspace{-150pt} \Xi_0 + (1 - \Xi_0)e^{1 - (1 + z)^{\nu}}\,.
\end{eqnarray}
The most relevant parameters are $\Xi_0$, which is the assymptotic value of $\Xi$, and the amplitude of the dark energy EOS deviation from -1, $A$. Figure \ref{fig:results} shows the resulting contours in the parameter space $\Theta = (\Xi_0, A, H_0, \Omega_{m0})$ when allowing for deviations of GR with the parametrizations. We also adapted existing codes for SNe and a combination of CMB and BAO data to include our MG parametrizations and used the results as priors. However, to exemplify, a particular model of the $\gamma$-gravity $f(R)$ theories \cite{ODwyer2013}, that is easily ruled out by other kinds of tests is represented with a green star: despite having $|f_{R0}|$ of order $10^{-2}$ it cannot be distinguished from $\Lambda$CDM.

\section{Conclusions and next steps}

\noindent

In this work, we studied the evolution of the GW distance in MG with special interest in $f(R)$-like theories. By generating a mock dataset of 1000 GWs emitted by BNS mergers as could be detected by the ET in near future, we investigated the constraints that the inference of the GW distances and redshifts of the sources could provide to cosmological and phenomenological parameters. Apart from measurements of the cosmological parameters, however, regarding $f(R)$ models, first, this test was shown to provide weak bounds to $f_{R0}$ when restricting to viable models; second, even when accounting for the modified background via $w_{\mathrm{DE}}$, we showed that it was not possible to distinguish between $\Lambda$CDM and typical $f(R)$ models that are already discarded with other probes. We also find that the GW bounds to the amplitude of the dark energy EOS $A$ are weaker than the ones from SNe and BAO/CMB data. Our conclusions indicate, therefore, that the test is not suitable to look for signatures of $f(R)$ modifications of gravity, neither from dark energy effects or modified friction in GW propagation, even in our optimistic scenario. Still, it constitutes an independent probe of gravity at cosmological scales that provides a measurement of deviations from GR at the percent level ($\Delta \Xi_0 \sim 10^{-2}$) and it could be quantitatively competitive for other theories within Horndeski and beyond.

Lastly, we remark that, as first highlighted in \cite{Saltas2014}, in our MG context, the same functions that modify the propagation of tensor modes also affect the scalar sector. For instance,
\begin{equation}
\Psi - (1 + \alpha_T)\Phi + (\alpha_M - \alpha_T)\mathcal{H}(\delta \phi/\dot{\phi}) = \Pi\,.
\end{equation}
In particular, the friction $\alpha_M$ makes the slip $\eta = \Phi/\Psi$ deviate from the GR value of unity, a feature usually generated whenever the matter content considered has anisotropic stress $\Pi$. Therefore, by combining slip measurements from LSS probes and GWs it might be possible to break the degeneracy between a scenario in which the gravity theory is GR with the presence of a matter sourced anisotropic stress, and a perfect fluid in MG. Studying the constraining power of the combination of future GW and LSS data on $\alpha_M$ is the subject of our current project.\footnote{in collaboration with M. Kunz and E. Bellini.}


\end{document}